\documentclass[cits]{PoS}
\usepackage{amsfonts,amssymb,latexsym,amscd,fontenc, times, mathptmx, graphicx}
\title{A new mechanism for dark matter generation from an interacting cosmological constant}
\ShortTitle{A new mechanism for dark matter generation...}
\author{Murli Manohar Verma\thanks{The author is thankful to  Frank Rieger and  Alexey Bogomazov  for their helpful  comments   and  acknowledges the  financial support  F. No. 37-431/2009 (SR)  received from  UGC, New  Delhi    towards this work.}\\
       Department of Physics,
       Lucknow University, Lucknow 226007, India\\
        E-mail: \email{sunilmmv@yahoo.com}}

\abstract{We propose  an alternative scenario for the dark matter generation  from an evolving cosmological constant which interacts with the dominant background in certain intermediate phase of the universe, and relaxes to the observed small value at present. In this way, it is shown that the interaction of the cosmological constant with the  radiation or matter might generate the dark matter densities with a varied mass spectrum in the universe with their characteristic arc-like frozen  signatures on the  Cosmic Microwave Background Radiation (CMBR). This approach also suggests a possible solution to the long standing cosmological constant problem.}

\FullConference{25th Texas Symposium on Relativistic Astrophysics - TEXAS 2010\\ 
December 06-10, 2010\\
Heidelberg, Germany}

\begin{document}

\section{Introduction} Dark energy has emerged as a major component of the universe following the observations of type Ia supernovae (SNe Ia) \cite{a1,b1} and the acoustic peak at the Legendre multipole $l_\emph{peak}\sim 197$ in the Cosmic Microwave Background Radiation (CMBR) power spectrum  \cite{c1}. While the SNe Ia luminosity-redshift data called for a component with negative pressure that could generate  required acceleration not explained  by the conventional model, the CMBR data required just to plug in the gap with positive energy density. These two diverse sets of  observations together needed  a specific equation of state (EOS) connected through the EOS  parameter $w$. The cosmological constant with $w_\lambda=-1$ in $\lambda$CDM   model seems to be the most viable candidate for this unknown component \cite{d1}. However,  this $\lambda$ component has been mostly  taken by researchers \cite{e1,f1,g1,h1,i1}  to be non-interacting with the background so that it maintains a constant energy density and yields an exponential expansion via the Friedmann equations as inflation in the early universe.

 Against this backdrop,  we have considered a different possibility of its interaction, since it seems hard to be convinced that a  presently dominant component  with density parameter $\approx 0.73$  that drives the dynamics of early inflation and the present acceleration in the universe must only stay sleepy with no  interaction with the co-existing  components of radiation or matter.

Thus we have shown previously \cite{j1} that  in a three-phase universe $\lambda$  must vary due to interaction  (then we call it a parameter) in the intermediate phase II known as  Q phase. In phase I and III, however,  it remains a constant and does not interact, where it causes  the required acceleration. Thus in Q phase we have  $T^{ik}_{(n){;k}}= -T^{ik}_{(\lambda){;k}}= Q^{i}$  and only the combined stress energy tensor is conserved. The interaction pulls down the energy density of the cosmological constant $\rho_{\lambda}$ quite fast, while the fall of matter $\rho_{n}$  occurs much more slowly  than without interaction.

In the  following  Section 2 we mention the background results for the $\lambda$-matter interaction and the evolution of their relative energy densities in Q phase.  In Section 3  we discuss  a scenario where  in this phase of the  universe the energy transfer from the decaying cosmological constant would generate the dark matter particles with wide mass spectrum whose signatures must be observed in form of the mass dependent arcs  on the CMBR power spectrum among other observable influences on structure formation.

\section {Decay of cosmological constant through interaction  in Q phase} The  model for interacting cosmological constant is in part motivated by the simple consideration of the effective curvature parameter $k_{\emph{eff}}=k- \lambda a^2/3$ in the Friedmann equations (with $c=1$)

\begin{eqnarray} \frac{2\ddot{a}}{a}+\frac{\dot{a}^2+k_{\emph{eff}}}{a^2}&=& 8\pi G T_{\mu}^\mu \label{p1}\end{eqnarray}
\begin{eqnarray}\frac{\dot{a}^2 + k_{\emph{eff}}}{a^2}&=&
\frac{8\pi G}{3}T_{0}^0\label{p2}\end{eqnarray}
 The CMBR observations suggest  $k_{\emph{eff}}=0$ that yields  $a^{-2}$ dependence of $\lambda$. With the time dependence $\lambda \propto H^{4/3}$ in matter dominated (MD) and $\lambda \propto H$  radiation dominated (RD) universe. Several authors have discussed forms of $\lambda$ dependent on $H$  \cite{p1,q1}.  With  $L_{\lambda}/L_{P}\sim 10^{61}$ ( where  $L_{\lambda}$ and $L_{P}$ are  the $\lambda$-  and the Planck length scales respectively), a plausible way to remove the inconsistency of $\lambda L_{P}^2 \sim 10^{-123}$ would be to find a  physical mechanism that allows a swift decline of $\lambda$. We have attempted to achieve this by means of the interaction between $\lambda$ and matter components.

 We mention in passing that our local observations, by virtue of their nature, would always indicate a spatially flat universe, and it is not clear how we relate it to the global geometry \cite{r1,t1}. Any motivation to  "solve the flatness problem" as claimed by inflationary theories appear to be actually silent, contrary to the common opinion, on the curvature parameter \emph{per se} since  the entire $k/a^2$ term is thrown out like "baby with the water" from  (\ref{p1}) and (\ref{p2})  because of inflating scale factor, and not because one gets $k=0$.(This term  might haunt again in the contracting phase of a closed universe !)

  In the phase I of the 3-phase model, the universe can undergo transition from RD to $\lambda$ dominated universe or remain in the $\lambda$  throughout this phase, depending on the initial constant value of $\rho_{\lambda}$. However,  in the next Q phase, we have energy conservation as
  \begin{eqnarray}\dot{\rho}_{n}+3H \rho_{n} (1+w_{n})= -\dot{\rho}_{\phi}-3H \rho_{\phi} (1+w_{\phi}) = Q\label{p3}\end{eqnarray} with  $\rho_{n}$, $w_{n}$ standing for dominant background and $\rho_{\phi}$, $w_{\phi}$ for the dark energy scalar field. The scalar fields with or without interaction have been discussed by several authors e.g.  \cite{k1,l1,m1} and their limits have also been noted \cite{n1}. However, we consider here an interacting cosmological constant. In $\lambda CDM$ model  $w_{\phi}=w_{\lambda}=-1$. Three alternative courses exist here. One,  $w_{\lambda}=-1$ is held fixed throughout the evolution of the universe with $\rho_{\lambda}=-\dot{Q}$, as we consider in this paper. Then it is found that interaction of $\lambda $ will reduce its energy density with the consequence of change in any one or more parameters of $\lambda, c $ or $G$. Two, if all these ($\lambda, c, G$) are truly constants of time,  $w_{\lambda}= -Q/3H \rho_{\lambda} -1$ and $Q=0$. Three, both $w_{\lambda}$ and $\rho_{\lambda}$ may vary like dynamical  quintessence.

  The scalar field dynamics is given as
  \begin{eqnarray}\ddot{\phi}+3H \dot{\phi} +V'(\phi) &=& - \frac{Q}{\dot{\phi}}\label{p4}\end{eqnarray}
  The solutions of (\ref{p3}) provide  $\rho_{n}\propto f a^{-3(1+w_{n})}$ with $f(\lambda/M_{P})=a^{Q/ H \rho_{n}}$ as the coupling function ($M_P= {(8\pi G)}^{-1/2}$ is the reduced Planck mass)  and keeping $Q/ H \rho_{n}$ constant,  while  $\rho_{\lambda}\propto  a^{-Q/ H \rho_{\lambda}}$. The interaction parameter clearly depends on the variation of the coupling function as
   \begin{eqnarray}Q \propto \dot{f}/a^{3(1+w_{n})}\label{p5}\end{eqnarray}

   In the absence of interaction $f=1$ and  $\rho_{n}$ declines as  $a^{-3(1+w_{n})}$  as expected. Regardless of the background being matter or radiation dominated, $f\propto Qt^{3}$ for constant $Q$. The fast increase of coupling function with time during Q phase is mainly responsible for conversion of energy of dark cosmological constant into dark matter particles.as also for slowing down  the rate of  fall of  $\rho_{n}$  in the concave-like manner. Thus with a non-zero $Q$  we have higher matter density at the end of Q phase than without interaction.

   For a spatially flat universe with $\Omega_{n}^0 +\Omega_{\lambda}^0=1$ we then have
  \begin{eqnarray}  \frac{Q}{\dot{\phi}}\sim H_{0}^2 M_{P}^2 (1+z)^{3(1+w_n)} f'(\phi)\label{p6}\end{eqnarray}
 \begin{eqnarray}  V'(\phi)= -\frac{Q}{\dot{\phi}}\label{p7}\end{eqnarray}
 It is found that for the observed range of $w_{\phi}\simeq -1$, ($w=-0.969\pm 0.061$ (stat)$\pm 0.065$ (sys)) \cite{d1} $\dot{\phi}\approx0$ and thus the right hand side of the (\ref{p4})  and (\ref{p7}) will be too high even for small value of $Q$. This shows the important role played by the interaction even if there is no cosmological constant proper with $w_{\lambda}=-1$, but instead,  a set of quintessence fields having $w_{\phi}\approx -1$. Thus the range of present observational uncertainties in determining EOS of dark energy is consistent with the concept of  interaction presented in this work.

 The coupling function shows the time dependence of the ratio of energy densities $r=\rho_{\lambda}/\rho_{n}$ as
 \begin{eqnarray}rf^{1+1/r}\propto a^4\label{p8}\end{eqnarray}  in RD  and  $\propto a^3$ in MD  universe. This gives
  \begin{eqnarray}\rho_{\lambda} \propto f^{-1/r}\label{p9}\end{eqnarray}.

  It is seen that the effects of interaction for fixed $w_{\lambda}$ are   masked with those of varying $w_{\lambda \emph{eff}}$ with no interaction. Thus

\begin{eqnarray} w_{\lambda \emph{eff}}= -1+\frac {1}{3r}\left(\frac{\ln f}{\ln a}\right). \label{p10}\end{eqnarray}

\section {Diagnostic for  coupling and dark matter generation} It was shown earlier that the third derivatives, and not the second derivatives,  of scale factor of the universe provide effective statefinders  \cite{u1}  for pinning down  the interaction among the components
\begin{eqnarray}u=\frac{\ddot{\alpha}}{aH^3}\label{p11}\end{eqnarray}
 \begin{eqnarray}s=\frac{u-1}{3(q-1/2)} \label{p12}\end{eqnarray}
 where  $\alpha=\dot{a} = aH$. In our case of cosmological constant this set becomes
\begin{eqnarray}u=1-\frac{9}{2}\left(\frac{Q}{3H(\rho_n + \rho_\lambda)}\right) \label{p13}\end{eqnarray}
 \begin{eqnarray}s=\frac{Q}{3H\rho_\lambda} \label{p14}\end{eqnarray}

 In terms of  (\ref{p14}) we have the coupling function as
 \begin{eqnarray}f= a^{3rs} \label{p15}\end{eqnarray}
 In non-interacting phases (initial phase I and the presently  ongoing phase III) no mechanism to generate dark matter can  operate and so $s=0$. However, such mechanism can clearly work in Q phase
with $s\neq 0$ (though not constant).
 In this way the universe has  greater matter density $\rho_{n}$ at the end of Q phase, thus alleviating the problem of producing light nuclei like deuterium-2 and lithium-7 in the  standard Friedmann-Robertson-Walker (FRW)  model. In addition to it,  the part of the decaying $\lambda$ energy is used up in generation of dark matter. These processes must have a deep influence on the structure formation in the subsequent course of evolution.

 We envisage a scenario  to generate  particles of dark matter (with Lorentz factor $\gamma$)  each with energy
 \begin{eqnarray}\xi= \int \Theta^{00} d\tau= \gamma mc^2  \label{p16}\end{eqnarray} where $\Theta^{00}$ is contributed by the ${T^{00}}$ of $ \lambda$ field. For the same given $\xi$ the more massive particles would travel more slowly than the lighter particles. Thus more massive particles would settle further away from the light cone envelopes, while the lighter particles traveling with relativistic speeds  remain closer  to the particle horizon. This particle mass distribution would in effect generate a multiple  arc-like  pattern on the Last Scattering Surface (LSS) around the recombination epoch ($z\approx 1000$) frozen on the CMBR,(and other observations  \cite{s1}),  each arc identifiable with a characteristic mass of definite dark matter particle (or its decay product releasing nett energy $\tilde{\xi}$). Equation  (\ref{p5}) shows the decay rate depends on the strength of interaction and as coupling grows  with time as $\propto t^3$ during Q phase greater is the amount of $\lambda$ energy used up to create the dark matter particles of broad mass spectrum.

 It is therefore expected to have an isotropic streaming of particles, produced  simultaneously (the vertex of the light cone lies at this  epoch  ) but  leaving their signatures in form of multiple arcs at LSS. For a  vertex at a  given epoch prior to the LSS, the arcs of smaller radii correspond to more massive particles  and those of large radius towards the horizon $R_H$ (horizon size $\approx$  0.18 Mpc at $t_{recombination}$ now blown to $\approx$ 200 Mpc) correspond to the lighter particles. The size of an  arc  $L\sim R_H v/c $  for a  typical particle having streaming velocity $v$  can  be related to its  mass  as

  \begin{eqnarray} m=\xi (1-L^2/R_H^2)^{-1/2}\label{p17}\end{eqnarray} Thus the mass density may be estimated from the vacuum considerations  \cite{o1}  with a suitable  cut-off  as
 \begin{eqnarray} \Pi= \frac {\hbar}{4\pi ^2} \int _0^{k_m} \omega_k k^2 dk (1-L^2/R_H^2)^{-1/2}\label{p18}\end{eqnarray}
 Putting  present energy scales $\sim 10^{-52} GeV^4$, $\rho_{\lambda}^0\simeq$ 4 GeV. this must include the uncertainty in mass given as
 \begin{eqnarray} \triangle m \geq \hbar/ \triangle L\label{p19}\end{eqnarray}
 where putting the maximum value of the present horizon $R_H^0 \approx 3000$ Mpc, we get $\triangle m\sim 10^{-32}$ eV.

 We expect the intensive dark matter search experiments, like XENON100, DAMA/LIBRA, CDMS II, as well as future LUX or  XMASS etc.  would  look for the lighter dark matter particles below $\sim10$  GeV  as the decay products of more massive particles,  and at the same time  for  the signatures of decay energy $\tilde{\xi}$  released in form of $\gamma$  radiation.

  However, the particle production being a continuous process throughout Q phase,  an overlapping of orders seems  inevitable for different sets of particles produced at different epochs. In this scenario it seems that the more massive particles would be the most suitable sites for the star formation in proto-galaxies due to stronger self-gravity, and so would yield age for the youngest stars. The above conclusions of the interacting cosmological constant  would  be accepted only at the touchstone of such  future observations.

\end{document}